\documentclass[a4paper,fleqn,usenatbib,useAMS]{mnras}
\usepackage{graphicx}	
\usepackage{amssymb}	
\usepackage{multicol}   
\usepackage{bm}		
\usepackage{amsmath}

\title[Photon Energy Evolution in TCAF scenario]{Temporal Evolution of Photon Energy emitted 
from Two Component Advective Flows: Origin of Time Lag}

\author[Arka Chatterjee, Sandip K. Chakrabarti \& Himadri Ghosh]
{Arka Chatterjee$^{1}$\thanks{E-mail: arka@csp.res.in}, Sandip K. Chakrabarti$^{2,1}$\thanks{E-mail: chakraba@bose.res.in} 
and Himadri Ghosh\thanks{himadri.ghosh@heritageit.edu}$^{3,1}$\\
$^{1}$Indian Centre for Space Physics, Chalantika 43, Garia Station Rd.,
             Kolkata, 700084, India\\
$^{2}$S. N. Bose National Centre for Basic Sciences, Salt Lake,
              Kolkata, 700098, India\\
$^{3}$Heritage Institute of Technology, Kolkata, 700107, India}
\date{Accepted XXX. Received YYY; in original form ZZZ}

\pubyear{2017}

\begin{document}
\label{firstpage}
\pagerange{\pageref{firstpage}--\pageref{lastpage}}
\maketitle

\begin{abstract}
X-Ray time lag of black hole candidates contains important information regarding 
the emission geometry. Recently, study of time lags from observational data revealed very intriguing 
properties. To investigate the real cause of this lag behavior with energy and spectral states, we  
study photon paths inside a Two Component Advective Flow (TCAF) which appears to be a satisfactory model 
to explain the spectral and timing properties. We employ the Monte-Carlo simulation technique to carry out the 
Comptonization process. We use a relativistic thick disk in Schwarzschild 
geometry as the CENtrifugal pressure supported BOundary Layer (CENBOL) which is
the Compton cloud. In TCAF, this is the post-shock region of the advective component. Keplerian disk 
on the equatorial plane which is truncated at the inner edge i.e., at the outer boundary of the CENBOL,
acts as the soft photon source. Ray-tracing code is employed to track the photons to a distantly 
located observer. We compute the cumulative time taken by a photon during Comptonization, 
reflection and following the curved geometry on the way to the observer. Time lags 
between various hard and soft bands have been calculated. We study the variation of 
time lags with accretion rates, CENBOL size and inclination angle. Time lags for different energy 
channels are plotted for different inclination angles. The general trend of variation of time 
lag with QPO frequency and energy as observed in satellite data is reproduced. 
\end{abstract}


\begingroup
\let\clearpage\relax
\endgroup
\newpage

\section{Introduction}

Many of the X-ray sources studied by UHURU, Ginga, Rossi X-ray Timing Explorer (RXTE) 
Monitor of All-sky X-ray Image (MAXI), NuSTAR and ASTROSAT are Black Hole Candidates (BHCs). 
The behavior of the spectral and timing properties across the time scales are of great interest 
to understand the geometry of the emitting gas. The spectrum mostly consists of a multi-color
black-body component at around $0.1-5$ keV and a power-law component 
which may go up to hundreds of keV often having an exponential cut-off. The study of the evolution in
the spectral state is considered to be the most promising way to learn about the intrinsic properties 
such as the mass $M$ and spin $a$ of black hole candidates and as well as the properties 
and dynamics of the flow accreting onto it. There are several models in the literature 
which attempt to understand these properties. Basically all the models start with the standard 
Shakura-Sunyaev (1973) Keplerian disk along with a hot electron cloud located nearby
(Takahara, Tsuruta \& Ichimaru, 1981; Maraschi, Roasio \& Treves, 1982; 
Sunyaev \& Titarchuk, 1989; Haardt \& Maraschi, 1991,1993).
In the present paper, we concentrate on the Two Component Advective Flow (hereafter TCAF) solution
(Chakrabarti \& Titarchuk 1995, hereafter CT95) whose every feature is derived from fundamental 
equations which govern the flow (Chakrabarti, 1990; 1996). This solution directly derives the 
physical flow parameters by fitting the
data with a minimal set of parameters (see, Debnath et al. 2014; Mondal et al. 2014; Jana et al. 
2015; Molla et al. 2016; Chatterjee et al. 2016) instead of just dwelling on the extrinsic 
and derivable properties such as the fluxes, spectral slopes and hard or softness 
as used in other models. The parameters required for 
TCAF fit of any spectrum are the accretion rates (halo rate $\dot{m_{h}}$ and disk 
rate $\dot{m_{d}}$), the shock location $X_{s}$ (which represents the size of Compton cloud) and the strength of shock $R$ 
(defined as the ratio of post-shock ($\rho_{+}$) and pre-shock density $\rho_{-}$) and of course 
the mass of the black hole $M$. 

For a set of flow parameters, TCAF provides the spectrum consisting of the multi-color black body and power-law components
with possible reflection and cut-offs combined. However, the photons at the observers frame
arrive after various physical processes, such as multiple scattering in the Compton 
cloud (i.e., the CENtrifugal pressure supported BOundary Layer or CENBOL in TCAF paradigm), 
reflections, focussing effects apart from the normal changes in time-delay due to space-time curvature at 
the emission region. Thus, hard and soft photons need not arrive simultaneously. 
The difference of time between the hard $H_t$ and the soft $S_t$ photons is called lag $t_{lag}$.
\begin{equation}
t_{lag}=H_t-S_t.  
\end{equation}
If $t_{lag}$ is positive, it is called the hard lag and the opposite is true for the soft lag.
The time lag is calculated by averaging the time of arrival of all photons for a given energy band. 
The observational confirmation of such hard and soft lags are regularly reported 
in the literature (e.g., Miyamoto et al. 1988; Cui et al. 2000; Qu et al. 2010; Uttley et al. 2011;
Dutta \& Chakrabarti, 2016). These lags are mostly studied for Low 
Frequency QPOs of Galactic black holes and perturbations in AGNs and Quasars. 
In this paper, we concentrate on Galactic black holes. It is believed that 
Comptonization plays a major role in the lag properties of Galactic black 
holes (Payne 1980). Numerous models (e.g., Cui et al. 1999; Nowak et al. 1999; Poutanen 2001) 
have been proposed to explain the lag properties. Ohkawa et al. (2005) explained the
lag behavior of GRS 1915+105 using Comptonization model. However, no unique 
photon emitting system which decides the lag behavior is theoretically 
established. Dutta \& Chakrabarti (2016, hereafter DC16) provided a list of  
contributors (e.g., Comptonization, disk reflection and gravitational lensing) to the
time lag and the possible outcome, attempting for the first time to explain 
difference in behavior of systems with both low and high inclination angles. 
Our present simulation is to actually reproduce magnitudes of these contributions and see
if any new effects are important. Compton upscattering naturally produces the hard lag. The connection 
with oscillation of Comptonization region to QPO frequency has already been 
established by Chakrabarti \& Manickam (2000, CM00) after extending the results 
of the numerical simulations of the oscillatory shocks in advective flows by 
Molteni, Sponholz \& Chakrabarti (1996) (hereafter MSC96). So, the lag variation  
with QPO frequency or shock location $X_s$ is the natural outcome of TCAF solution. 
Lags are generally produced due to the light crossing time through different regions of the accretion disk
and thus the main challenge is to find out how the path lengths of the observed photons vary with their energies.

Spectral variabilities with inclination angle is observed and reported several times in the literature (e.g., Motta et al. 2015; 
Heil et al. 2015). Since it is impossible for us to observe a particular BHC in different inclination
angles, we must carry out the Monte-Carlo simulation of a realistic system to check if the lag properties are 
generally different from various angles as observed. Using the same procedure, earlier Ghosh et al. (2011) 
established that the spectrum becomes harder if observed at a high inclination angle. However, that was 
purely due to projection effects. Chatterjee, Chakrabarti \& Ghosh (2017, hereafter CCG217) 
performed ray-tracing of the photons emitted by a Monte-Carlo method and projected them on the observer plane. 
They showed that the image and the spectra both change with the inclination angle as the contribution 
of hard photons increases. This motivates us to study the variation of time lag with 
various flow parameters, such as, the accretion rate, CENBOL size (shock location), photon energy, theoretically
calculated QPO frequency from the shock location and, of course, the inclination angle.     

In this paper, we present simulated results of time lags found in the phasor part of the 
Power Density Spectrum (PDS) of BHCs using TCAF geometry. In the next Section, we
describe the geometry and thermodynamical properties of TCAF used in the simulation. 
In \S 3, we briefly describe the ray-tracing mechanism. Section 4 describes the Monte-Carlo simulation 
process. We then present our results in Sec 5. This is followed by Discussions
and Conclusions.    

\section{Geometry of the Problem}

TCAF mainly consists of a Keplerian flow with higher angular momentum and an advective, sub-Keplerian component
with low angular momentum surrounding the Keplerian disk. Sub-Keplerian flow or the halo component 
puffs up after reaching the centrifugal barrier known as the Shock Location ($X_s$). Depending 
on the criteria for a shock to form, i.e.,
whether the Rankine-Hugoniot conditions are satisfied or not, the shock may be stationary (Landau \& Lifshitz, 1959) 
or oscillating depending upon the geometry and the thermodynamical properties of flow parameters. 
In Molteni et al. (1994), it was pointed out that the post-shock region or 
CENBOL behaves like thick disks  as envisaged by Abramowicz et al., 1978; Kozlowski et al., 1978; 
Paczy\'nski \& Wiita, 1980; Begelman, Blandford \& Rees, 1982; Chakrabarti, 1985; hereafter C85) though the 
simulation results are more accurate due to the presence of radial velocity.
In the present paper, we consider C85 prescription to obtain relativistic 
thick disk parameters in our geometry. This hot region
acts as the Compton cloud where soft photons from the Keplerian disk are intercepted and inverse-Comptonized
to create a power-law tail as theoretically calculated in CT95. Below, we briefly 
describe the geometry and thermodynamical properties of the CENBOL 
and the Keplerian disk. We consider the natural units where $r_g=\frac{2GM}{c^2}=2$ with $G=1$,~$M=1$ 
and $c=1$ throughout the paper. Accretion rate is measured in Eddington unit, i.e., 
$\dot{m}_d=\frac{\dot {M}}{\dot {M_{Edd}}}$. In this work, we considered the outer edge of truncated 
Keplerian disk to be extended up to $100 r_g$. CENBOL size varies from $20 r_g$ to $65 r_g$. Outer
edge of CENBOL is the inner edge of the Keplerian disk. 

\subsection{Compton Cloud or CENBOL}
Euler's equation for a perfect fluid can be written as,
\begin{equation}
\frac{\nabla p}{p+\epsilon} = -\ln(u_{t}) + \frac{\Omega \nabla l}{(1-\Omega l)}
\end{equation}
(for details, see, C85). Also, $l=-u_{\phi}/u_{t}$ is the specific angular momentum 
and $\Omega =u^{\phi}/u^{t}$ is the relativistic angular
velocity. Four velocity is considered as $[u_{t},0,0,u_{\phi}]$. To enforce equipotential 
surfaces to remain the same as the equipressure surface, we choose the barotropic ($p(\epsilon)$) equation of state. 
So, Euler's equation becomes,
\begin{equation}
W-W_{in} = \int_0^p\frac{dp}{p+\epsilon}=\int_{{u_{t}}_{in}}^{u_{t}} \ln(u_{t}) - \int_{l_{in}}^{l}\frac{\Omega \nabla l}{(1-\Omega l)},
\end{equation}
where, the natural angular momentum distribution ($l=c\lambda^n$, $c$ and $n$ are constants) yields the 
von-Zeipel relation $\Omega=\Omega(l)=c^{2/n}l^{1-2/n}$ with  
$\lambda=\frac{r\mathrm{sin}\theta}{(1-\frac{2}{r})^{1/2}}$ (C85, Paper1). Since, by definition, 
the angular momentum
distribution of  a thick disk intercepts the Keplerian distribution at two points, namely,
the inner boundary ($r_{in}$) and the centre of disk ($r_c$), it is easy to calculate the values of $c$ and $n$ using algebraic equations 
from those points (C85). Once, we fix the constants of the flow, the equipotential surfaces filled with
matter creates the CENBOL region. In reality, density drop near the outer edge is faster than what is chosen here. This is 
due to the presence of a shock front. The gradual drop of density at the outer edge of the CENBOL
which we have chosen here would correspond to the presence of a weaker shock.

For Comptonization, we need to provide the exact number density and temperature of electron in each grid of the
CENBOL region. We use polytropic equation of state, $p=K\rho^{\gamma}$, where $p$ is pressure, 
$\rho$ is the matter density, $K$ is the constant or measure of entropy of the system and $\gamma$ 
is the polytropic index. Value of $\gamma$ is considered to be $4/3$. Typical electron number density  
and temperature is shown in Figure 1. 

\begin{figure}
\centering
\vbox{
\includegraphics[width=0.45\columnwidth]{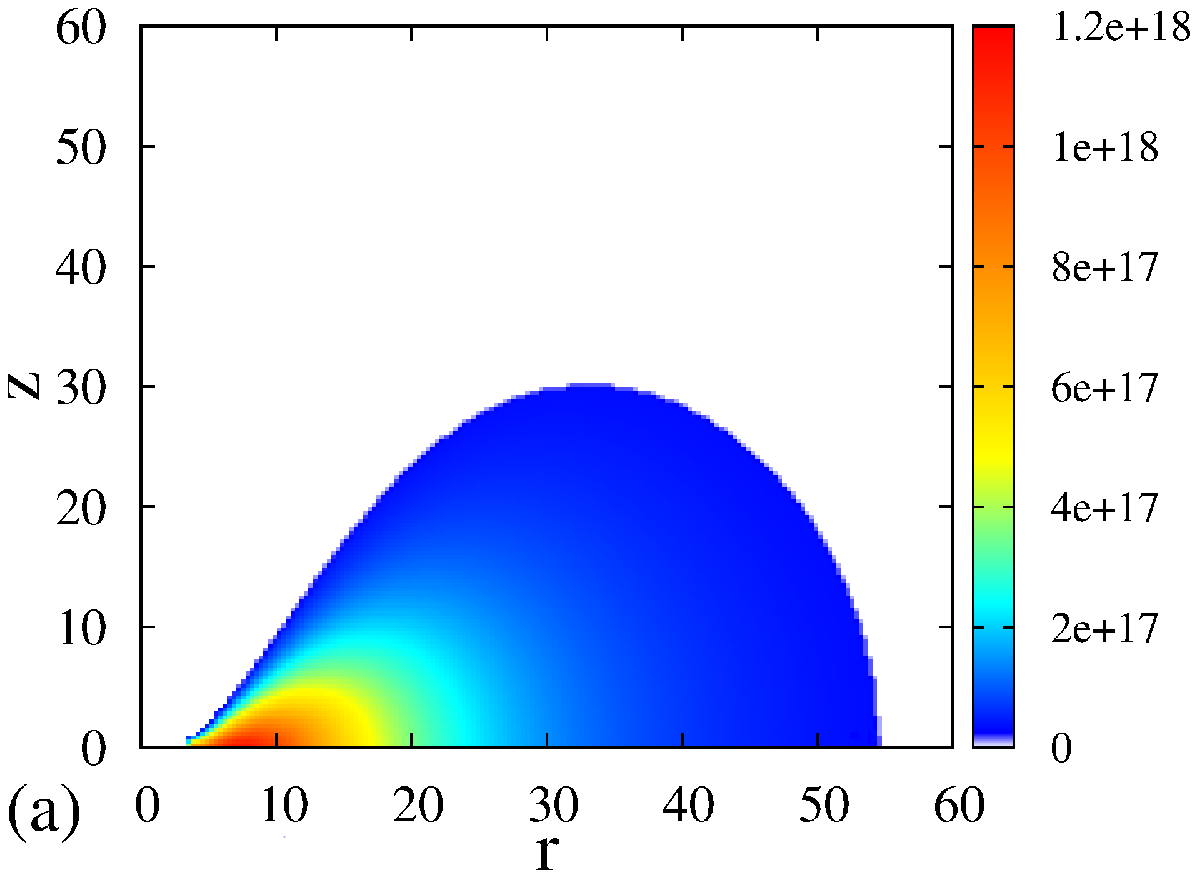}
\includegraphics[width=0.45\columnwidth]{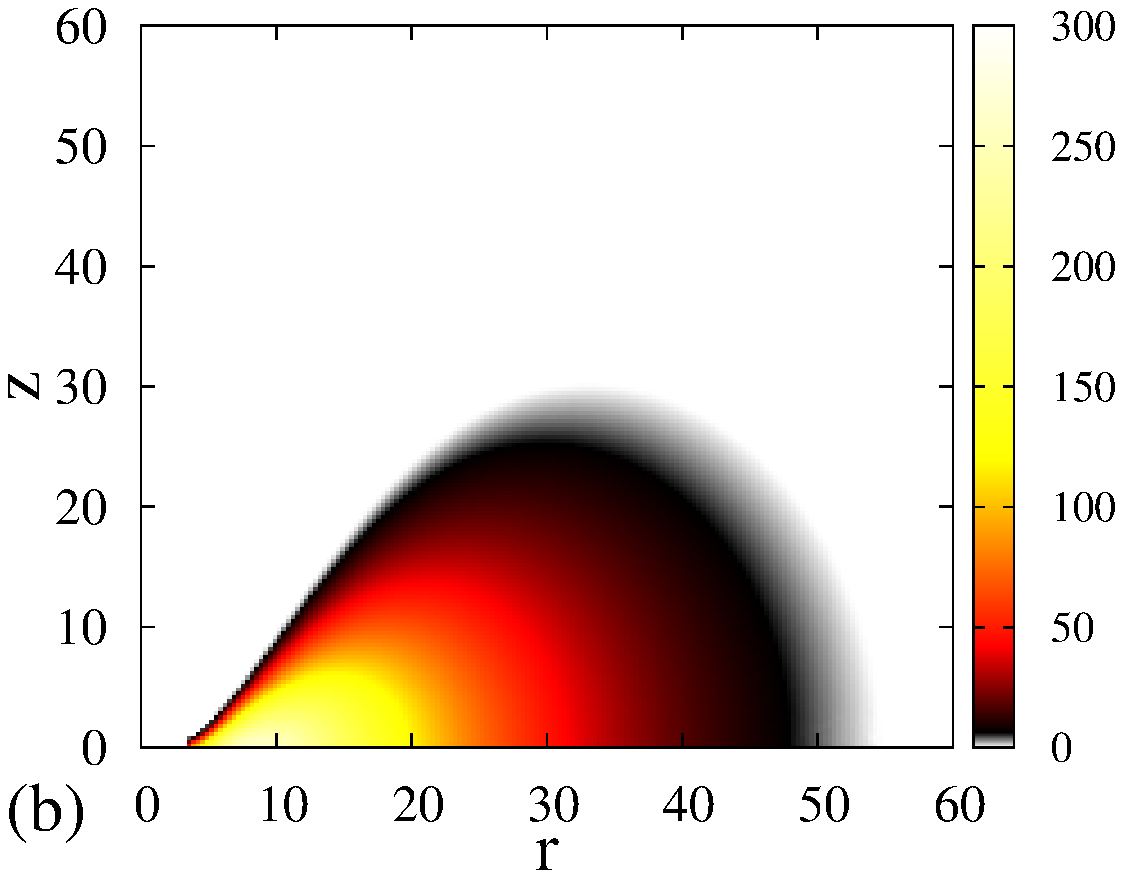}}
\caption{Typical (a) electron number density (in per $cm^3$) and (b) temperature (in keV)
distributions are shown. The central region of CENBOL contains the maximum density and 
temperature.}
\end{figure}

\subsection{Keplerian disk as the Soft Photon Source}

Keplerian disk is situated outside the Compton cloud boundary or the shock location $X_s$ on the equatorial plane. Theoretically,
it has a higher viscosity (see, Chakrabarti 1990; Chakrabarti 1996; Chakrabarti \& Das 2004 and 
Nagarkoti \& Chakrabarti 2016) and much lower temperature than the CENBOL region. Thermal 
photons are produced in the Keplerian disk via black body emission. The truncated Keplerian 
disk extends up to $100~r_g$. Radiation profile of Keplerian disk adopted from Page \& Thorne, 
(1974). Inner edge of the Keplerian disk and outer edge of the CENBOL coincides. Flux and 
temperature profiles are given below. 

\begin{equation}
\begin{aligned}
F(r) = 
\frac{F_c(\dot{m}_d)}{(r-3)r^{5/2}}                                 \\
\times \bigg[\sqrt{r}-\sqrt{6}+
\frac{\sqrt{3}}{3}log\bigg(\frac{(\sqrt{r}+\sqrt{3})(\sqrt{6}-\sqrt{3})}
{(\sqrt{r}-\sqrt{3})(\sqrt{6}+\sqrt{3})}\bigg)\bigg] 
 \\ and \\                                  
T(r)= \bigg(\frac{F(r)}{\sigma}\bigg)^{1/4}
\end{aligned}
\end{equation}
where, $F_c(\dot{m}_d)=\frac{3m\dot{m}_d}{8\pi r_{g}^3}$, with $\dot{m}_d$ being 
the disk accretion rate in Eddington unit, $\sigma=\frac{2\pi^5k^4}{15h^3c^3}$ is 
the Stefan-Boltzmann constant.

Photon flux emitted from the Keplerian disk surface having radius $r$ to
$r+\delta r$ is given by,
\begin{equation}
n_{\gamma}(r)=\bigg[\frac{4\pi}{c^2} \bigg(\frac{k_bT(r)}{h}\bigg)^3 \times \zeta (3)\bigg]~cm^{-2}s^{-1},
\end{equation}
where, $\zeta (3)=\sum^\infty_1{l}^{-3} = 1.202$ is the Riemann zeta function.
The rate of photons emitted Keplerian disk having radius $r$ to $r+\delta r$ is given by
\begin{equation}
dN(r) = 4\pi r\delta rn_{\gamma}(r)~s^{-1},
\end{equation}
(see, Garain, Ghosh \& Chakrabarti, 2014, for further details).

For the sake of simulations of photon injection to CENBOL, the 
Keplerian disk is divided into several annuli of width $D(r) = 0.1$. $T(r)$ is 
the temperature of the mean radius of each annulus. The total number of photons turn out
to be $\sim 10^{39}$ --$10^{40}$ per second for $\dot{m}_d=0.1$. To save 
computational time, we bundle photons using a weightage factor $f_W=\frac{dN(r)}{N_{comp(r)}}$
where $N_{comp}(r) = 10^8$.

We calculate soft photon energy using Planck distribution law for $T(r)$. 
The number density of photons ($n_\gamma(E)$) which correspond to an energy $E$ is given by,
\begin{equation}
n_\gamma(E) = \frac{1}{2 \zeta(3)} b^{3} E^{2}(e^{bE} -1 )^{-1}, 
\end{equation}
where, $b = 1/kT_{k}^{disk}(r)$ (e.g., Ghosh, Chakrabarti, Laurent, 2009, hereafter GCL09).

\section{Ray Tracing Process }

The ray-tracing equations are derived from field equation using non-vanishing Christoffel terms 
in Schwarzschild geometry,
\noindent
\begin{equation}
\frac{d^2x^{\mu}}{dp^2}+ {\Gamma}_{\nu\lambda}^{\mu}\frac{dx^{\nu}}{dp}\frac{dx^{\lambda}}{dp} = 0,
\end{equation}
where, ${\mu} = [0,1,2,3]$; $x^{0} = t$, $x^{1} = r$, $x^{2} = \theta$ and $x^{3} = \phi$,
$p$ is the affine parameter. We obtain four coupled, second order differential equations 
for photons where 4-momentum $P_t=E=(1-\frac{2}{r})\frac{dt}{dp}$ and 
$P_{\phi}=L=r^{2}\mathrm{sin}^{2}\theta\frac{d\phi}{dp}$ definitions are used
(Chandrasekhar, 1983) to shift the base from an affine parameter $p$ to time $t$. 
Thus, we retain three equations in $3D$ geometry with time as the base.
\begin{equation}
\begin{aligned}
\frac{d^2r}{dt^2} + \frac{3}{r(r-2)}\bigg(\frac{dr}{dt}\bigg)^2 - 
(r-2)\bigg(\frac{d\theta}{dt}\bigg)^2 - \\ (r-2)r\mathrm{sin}^2{\theta}\bigg(\frac{d\phi}{dt}\bigg)^2 
+ \frac{r-2}{r^3} = 0,\\
\frac{d^2\theta}{dt^2} + \frac{2r-6}{r(r-2)}\bigg(\frac{d\theta}{dt}\bigg)\bigg(\frac{dr}{dt}\bigg) 
- \mathrm{sin}{\theta}\mathrm{cos}{\theta}\bigg(\frac{d\phi}{dt}\bigg)^2 = 0 ~\mathrm{and}\\
\frac{d^2\phi}{dt^2} +\frac{2r-6}{r(r-2)}\bigg(\frac{d\theta}{dt}\bigg)\bigg(\frac{dr}{dt}\bigg)
 + 2\mathrm{cot}{\theta}\bigg(\frac{d\theta}{dt}\bigg)\bigg(\frac{d\phi}{dt}\bigg) = 0.
\end{aligned}
\label{eq:xdef}
\end{equation}

Velocity components are derived from the transformation equations (Park, 2006) and written as, 
\begin{equation}
\begin{aligned}
v^{\hat{r}}=\frac{d\hat{r}}{dt}=\frac{r}{(r-2)}\frac{dr}{dt},
~v^{\hat{\theta}}=\frac{d\hat{\theta}}{dt}=\frac{r\sqrt{r}}{\sqrt{(r-2)}}\frac{d\theta}{dt}\\ 
\mathrm{and} ~
v^{\hat{\phi}}=\frac{d\hat{\phi}}{dt}=\frac{r\sqrt{r}\mathrm{sin}{\theta}}{\sqrt{(r-2)}}\frac{d\theta}{dt}.
\end{aligned}
\label{eq:xdef}
\end{equation}
The process of Ray-Tracing is similar to CCG17.

\section{Monte-Carlo Comptonization}

\begin{figure}
\centering
\vbox{\includegraphics[width=1.0\columnwidth]{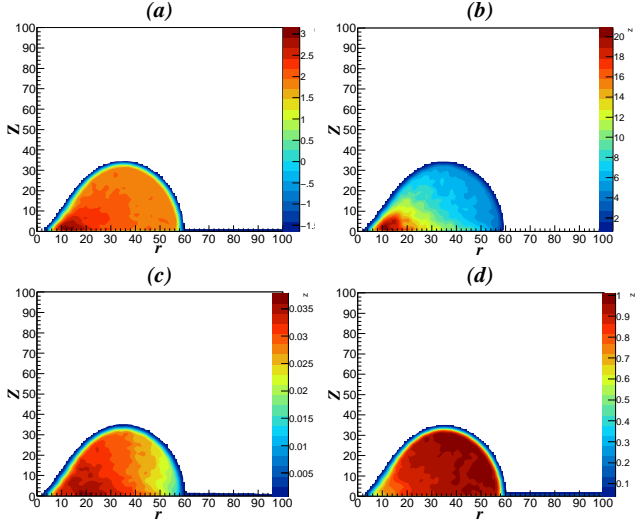}}
\caption{(a) Energy ({\it in keV plotted in $log$ scale}) of escaping Photons, (b) number of scattering          
({\it dimensionless number}) suffered by emergent photons from the CENBOL, (c) contours of constant time ({\it in sec}) 
of emission of photons after Comptonization and (d) vertical component of velocity of photons ({\it in the 
unit of velocity of light}) are represented for a CENBOL which has the outer edge at around $60r_g$.}
\end{figure}

Using straight line path of photons between two successive scatterings inside the Compton cloud 
saves some computational time without any loss of physics (Laurent \& Titarchuk, 1999). The 
injected photons are generated from the Keplerian disk at a random direction which required use 
of three random numbers. In the Monte-Carlo simulation, the functional form of $\mathrm{cos}\theta$ 
for the flux dependence of photons  on $\theta$  is implemented as is the case where 
temperature gradient inside the Keplerian disk is along Z axis. Inside the CENBOL, a critical 
optical depth $\tau_c$ is set up from a random number after each scattering. When the traversed 
optical depth of a given photon crosses $\tau_c$, it is allowed to make the next scattering. 

The Klein-Nishina scattering cross section $\sigma$ is given by:
\begin{equation}
\sigma = \frac{2\pi r_{e}^{2}}{x}\left[ \left( 1 - \frac{4}{x} - \frac{8}{x^2} \right) 
ln\left( 1 + x \right) + \frac{1}{2} + \frac{8}{x} - \frac{1}{2\left( 1 + x \right)^2} \right],
\end{equation}
where, $x$ is given by,
\begin{equation}
x = \frac{2E}{m c^2} \gamma \left(1 - \mu \frac{v}{c} \right),
\end{equation}
$r_{e} = e^2/mc^2$ is the classical electron radius and $m$ is the mass of the electron.

The energy exchange via Comptonization or inverse Comptonization is calculated in each step assuming
an electron with randomly chosen momentum vectors. While travelling from one scattering centre
to another, the gravitational redshift acts on a photon changing its frequency.
The process continues until the photon leaves CENBOL region or is sucked in by the
black hole. The process is similar to what was used in GCL09; Ghosh, Garain, Chakrabarti, 
Laurent, 2010; Ghosh et al. 2011 (hereafter GGGC11); CCG17.
 
Figure 2 is the outcome of our Monte-Carlo simulated Comptonization code. The last scattering point
of photons are stored. The energy contours (in Panel a) shows the
most energetic photons came from the central region where they suffered most scatterings.
Panel (b) shows the color gradient map of number of scatterings suffered by individual photons.
Contours of constant time of emission
are represented in Panel (c). The more time a photon spends inside the CENBOL the
more scattering it suffers. Panel (d) shows the magnitude of velocity along the z-axis. It is
to be noted that the azimuthal component of velocity ($v_z$) is nearly unity in upper side of
CENBOL. This means the photons from this region will go along the z-axis.

\section{Results}
The results shown here assume the mass of the black hole to be $10$ solar mass.
\subsection{Spectra}

We used Monte-Carlo method to simulate Comptonization inside the CENBOL region. Figure 3a shows the 
injected soft photon spectrum. Composite emission spectra for 
disk  and scattered Comptonized photons with the variation of CENBOL size are shown in Fig. 3b. 
The hardest spectrum corresponds to the largest CENBOL region. The black-body peak 
of the spectra shifts in lower energy range with the increasing CENBOL size since the inner edge of the
Keplerian disk recedes from the black hole.

\begin{figure}
\centering
\vbox{
\includegraphics[width=1.0\columnwidth]{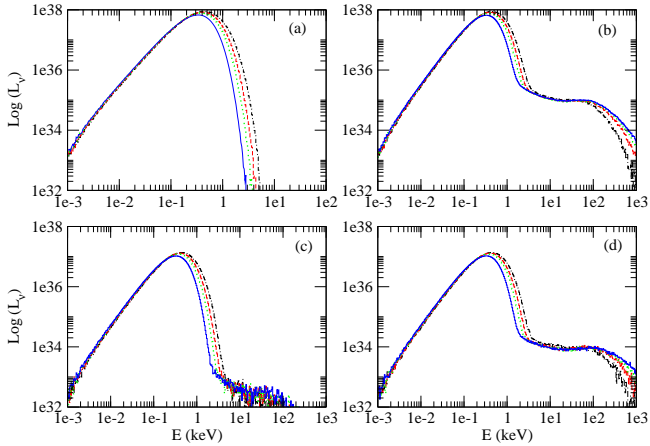}}
\caption{Spectral variation with the outer edge of CENBOL located at $20r_g$ (solid-black), 
$30r_g$ (dash-dash-dot-red), $40r_g$ (dashed-green) \& $65r_g$ (dot-dot-dash-blue). (a) 
the injected spectra, (b) Emergent Comptonized spectra for photons emitted in all directions, (c) spectra as
seen from nearly along the Z-axis and (d) spectra seen nearly along the equatorial plane.}
\end{figure}

These results agree with the spectral shape discussed in CT95 and Chakrabarti (1997).
Fig. 3(c-d) represent the same spectra for the lowest ($0^\circ-9^\circ$) 
and the highest ($81^\circ-90^\circ$) inclinations respectively. The hardness 
of the source spectrum over the inclination 
angle is profoundly visible in these two panels. Detailed information on inclination 
dependent spectra was provided in Ghosh et al. (2011) and CCG2017.  
 
\subsection{Time lag with CENBOL size}
It has been reported in DC16 that the time lag increases with 
the increasing of the CENBOL size i.e., the shock location $X_s$.

\begin{figure}
\centering
\vbox{
\includegraphics[width=1.0\columnwidth]{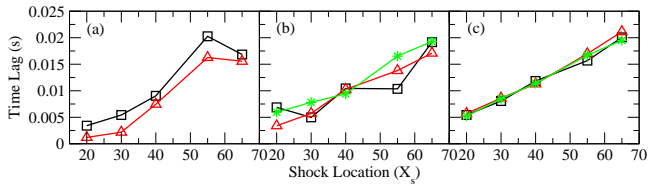}}
\caption{Time lag ($in~sec$) between hard band ($5-100$) keV and soft band
($0.1-5$) keV is plotted against Shock Location ($X_s$).
(a) Black (square) and red (triangle) lines represent $0^{\circ}$ \& $10^{\circ}$, 
(b) Black (square), red (triangle) \& green (star) lines
represents $20^{\circ}$, $36^{\circ}$ \& $50^{\circ}$ respectively and (c) Black (square),
red (triangle) \& green (star) lines represent $60^{\circ}$, $70^{\circ}$ \& $80^{\circ}$ respectively. }
\end{figure}
Following this, we simulate cases with variable CENBOL size. The hard band is considered
$5$ to $100$ keV and soft band is considered as $0.1$ to $5$ keV as in XMM satellite. 
From our simulations, we see that the time lag actually increases with the 
shock location. Figure 4 is plotted for three different bins of 
inclination angle: (a) low inclination ($0^{\circ}$ \& $10^{\circ}$), 
(b) mid range inclination ($20^{\circ}$, $36^{\circ}$ \& $50^{\circ}$) and (c) for 
high inclination angles ($60^{\circ}$, $70^{\circ}$ \& $80^{\circ}$). For panel (c), we 
see the monotonical increase of time lag w.r.t CENBOL size. This region is mostly
dominated by the lensing and disk reflection. Hard photons that have generated
well inside the Compton cloud takes time to come out. Note that the time lag 
we obtained is slightly smaller than generally observed values. This is because we 
use small CENBOL sizes to save computational time. Further details are provided in 
Sec. 5.5.  

\subsection{Time Lag with accretion rate}
We study accretion rate variation for a fixed CENBOL size to see how it affects the
lag features. The results are plotted in Fig. 5 for four different inclination angles. Accretion
rates of the Keplerian component are written in each panel. 
Apart from very low inclination angle we did not see any soft lag. 
We see the marginal trend of higher accretion rate causing higher hard lag. 
From CT95, it is known that higher $\dot{m}_d$ causes softer spectrum for a given sub-Keplerian
Component rate (which, together with the Keplerian rate, determines the CENBOL optical depth). 
The soft photons intercepted by the Compton cloud spends 
more time and are more energized by CENBOL. So, the value of hard lag increases in for
higher energy bins. This is the contribution of Comptonization to the lag feature.
\begin{figure}
\centering
\vbox{
\includegraphics[width=1.0\columnwidth]{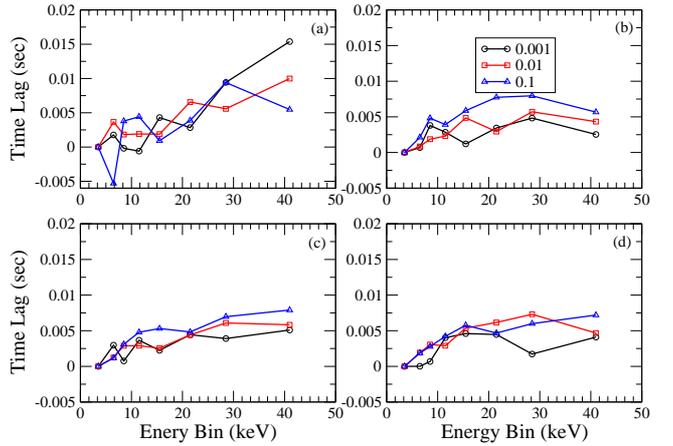}}
\caption{Time lag w.r.t $2.0-5.0$ keV plotted for three accretion rates (marked) as a 
function of observed energy at $0^{\circ}$, $36^{\circ}$, $60^{\circ}$ \& $80^{\circ}$ 
inclination angles in (a), (b), (c) \& (d) respectively. CENBOL size is kept fixed at $65 r_g$.}
\end{figure}

The change in $t_{lag}$ for each energy bin for different $\dot{m}_d$ fluctuates. 
However, the lag magnitude generally decreases with the inclination 
angle. Soft lag in Fig. 5(a) is found for accretion rate $0.1$ where the 
observer is face on with the disk. From Panel (d) of Fig. 2, we get the 
idea of vertical velocity distribution of photons. The top of the CENBOL 
contains photons having $v_z=1$. Those reach the observer earlier than the 
soft photons from Keplerian disk. Since they have scattered from the outer 
edge of the CENBOL, they are much likely to be $\leq 10$ keV. They will 
contribute to the soft lag for low inclination angle sources. It is to be noted 
that to understand the complete variation of time lag with accretion rate
one needs to supply two types of accretion rates (halo rate ($\dot{m}_h$) and disk 
rate $\dot{m}_d$) (CT95). The hydrodynamical simulations using two accretion rates
are out of the scope of this paper and will be reported elsewhere.

\subsection{Variation of Lag with Energy of Photon}

For Galactic black hole candidates, Comptonization is an essential component to produce
the hard photons. Mostly, inverse Comptonization in the CENBOL region generates the harder
photons. Most energetic photons are ejected from the central region (Fig. 2a) of CENBOL 
where electron density and temperature are maximum. Those photons suffer maximum number of 
scatterings as depicted in Fig. 2b. The more time photons spend inside the CENBOL the probability
of hardening becomes higher. So, Comptonization in general provides hard lag which can be
seen in Fig. 6. From Fig. 6, we find that the lag increases with decreasing    
\begin{figure}
\centering
\vbox{
\includegraphics[width=1.0\columnwidth]{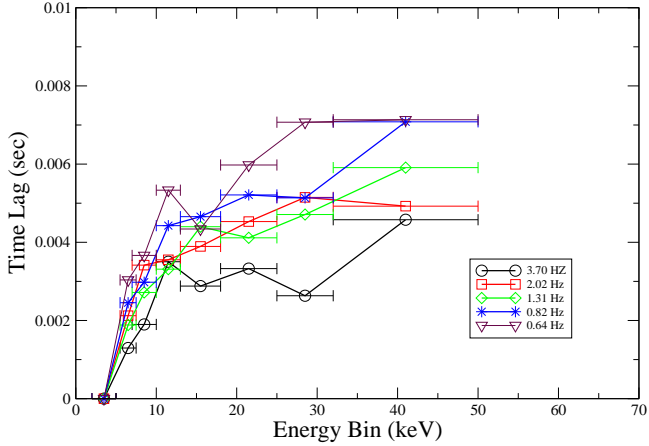}}
\caption{Time lag w.r.t. $2.0-5.0$ keV is plotted for various energy bins at 
a range of theoretically obtained QPO frequencies for a fixed inclination 
angle of  $70^{\circ}$.}
\vskip 0.30cm
\end{figure}
QPO frequency. These QPO frequencies are theoretically obtained from the relation
provided by CM00 and is given by
\begin{equation}
\nu_{qpo}=\frac{C}{RX_s(X_s-1)^{1/2}}\sim \frac{C}{RX_s^{3/2}},
\end{equation}
where, $C$ is a constant. $R$ \& $X_s$ are compression ratio and shock location 
respectively. MSC96 has showed that the time period of QPO oscillation is comparable
with the infall timescale of the postshock region ($t_{infall}\propto X_s^{3/2}$). Thus,
the QPO frequency will be inversely proportional to $X_s^{3/2}$. This relation has
predicted QPO frequencies for GRS 1915+105 (CM00) and for many other sources (Debnath, 
Chakrabarti \& Mondal, 2014). Observational evidence of such lag behavior is given 
in DC16. The lag magnitude is somewhat lesser than what people had observed (Nowak et al. 
1999; DC16 and Misra et al. 2017). This could be because of the smaller size of the 
accretion disk and Compton cloud we have considered 
for our simulation. To reduce computational time, we have restricted ourselves 
within $100r_g$ of the Standard disk. Bigger CENBOL size ($\sim 500 r_g$) would
provide an energy dependent lag up to $0.03$ Sec. A general trend of dipping in the curve 
can be seen at $13-17$ keV energy band in Fig. 6. These are the photons which suffered
more scatterings, but, from the outer layer of CENBOL than that by the photons in $7-13$
keV energy band. So, they arrive a little earlier.

\subsection{Variation of Lag with inclination angle}
In our previous studies (Chatterjee \& Chakrabarti 2015a, 2015b), 
we have considered only lensing effect by using pure Keplerian disk. There, 
we have shown the time of arrival changes drastically with increasing 
inclination angles.
\begin{figure}
\centering
\vbox{
\includegraphics[width=1.0\columnwidth]{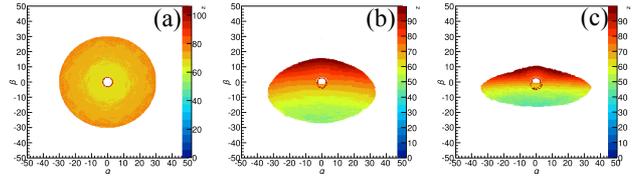}}
\caption{Time of arrival (in $r_g/c$ unit) of photons are mapped on top the images Keplerian 
disk geometry only. Colorbar represents the time of arrival to a distant observer. Inclinations 
angles of the observer are $0^{\circ}$, $50^{\circ}$ \& $70^{\circ}$ in panels (a), (b) \& 
(c) respectively. Note the upper band of the color-bar where all the secondary photons 
have arrived much later than the primary photons. All secondary photons are considered 
to produce this image.}
\end{figure}
Fig. 7 is a time of arrival map of Keplerian disk. 
For $0^\circ$ case, photons from inner region of the disk come earlier than the outer 
region. But, the secondary photons (those which encircle the black hole once or more)
originated from various locations of disk may take much longer time of arrival as depicted 
in Fig. 7a (points beyond $80$ in the z-range). At $50^\circ$ , the same disk looks 
different. Time of arrival of photons drastically changes at $70^\circ$. Light crossing 
time splits two sides of the Keplerian disk. The secondary photons take even more 
time to reach the observer frame as the observer moves to higher inclination angles. Here,
the inner boundary of Keplerian disk is at $6~r_g$ while the outer boundary is at $30~r_g$.
The position of the observer on azimuthal plane is at $45^\circ$. From this Figure, 
one can easily estimate the light crossing time for much larger accretion disks and for 
various observer inclinations. For the purpose of observational data, we have added Comptonization
and disk reflection into the study. After addition, the composite nature of time lag 
with the variation of inclination angle has been depicted.

DC16 explained the hard and soft lag features using XTE J1550-564 
($\sim 75^\circ$ inclination) and GX 339-4 ($\sim 50^\circ$ inclination). The processes 
that take part in the time lag are Comptonization, disk reflection and gravitational lensing.
In Fig. 8, we show the lag behavior of $5-7$ keV and $25-30$ keV band photons
with increasing CENBOL size. The separation between these two lags increases with 
increasing CENBOL size. Photons generated via inverse Compton scattering can come 
out from any optical depth of the Compton cloud (see, Fig. 2 and CCG17 for image 
description). Relatively low energy photons, such as $5-7$ keV bin are mostly produced 
from the outer layer of CENBOL and $25-30$ keV or even higher energy 
photons come from inner regions. So, with increasing size of CENBOL the distance between
outer layer and central region increases and time taken by a harder photon increases.
\begin{figure}
\centering
\vbox{
\includegraphics[width=1.0\columnwidth]{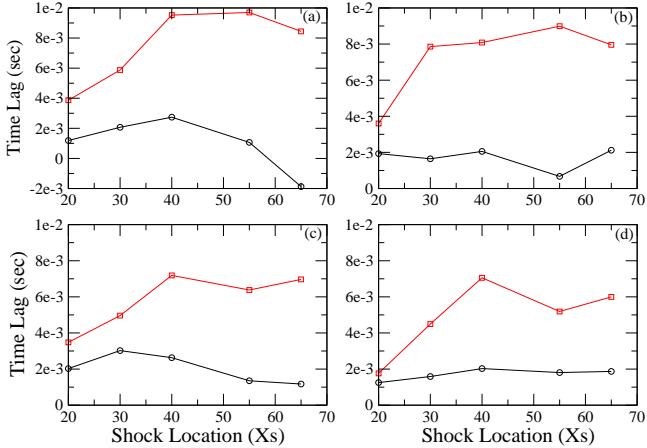}}
\caption{Time lags of $5-7$ keV (solid-black-circle) and $25-30$ keV 
(solid-red-triangle) photons with respect to soft band $2-5$ keV is plotted 
for $10^{\circ}$, $36^{\circ}$, $60^{\circ}$ \& $80^{\circ}$ in panels (a), (b), 
(c) \& (d) respectively.}
\end{figure}

In Fig. 8, we see the lag variation of $5-7$ keV photons in first three panels. 
In panel (d), for $80^{\circ}$ inclination angle, the lag variation with respect to 
CENBOL size for $5-7$ keV is negligible. This is due to the predominant edge view of 
CENBOL from that angle. Keplerian disk is extended at the edge 
of Compton cloud. Soft photons from Keplerian disk participate in the reference 
energy band of $2-5$ keV. So, the lag between the reference band and 
$5-7$ keV band still remains almost constant for variable Compton cloud in this near 
edge on scenario. It is to be noted that, for Fig. 4c, the hard energy band 
is taken from 5-100 keV. For Fig. 8d, the energy bands are splitted (5-7,
7-10, 10-13, 13-18, 18-25 keV etc.). Fig. 8 as a whole is meant to show the lag 
properties of each separate bands. We can see the monotonic increase (Fig. 4c) in 
the lag properties for cumulative bands as it is showing a macroscopic view. 
In order to show that the nature is not exactly same for each individual bands, we have plotted Fig. 8.
Non-monotonicity of the lag in Fig. 8 could be attributed to the non-monotonicity of 
temperature or density inside the CENBOL. 

\begin{figure}
\centering
\vbox{
\includegraphics[width=1.0\columnwidth]{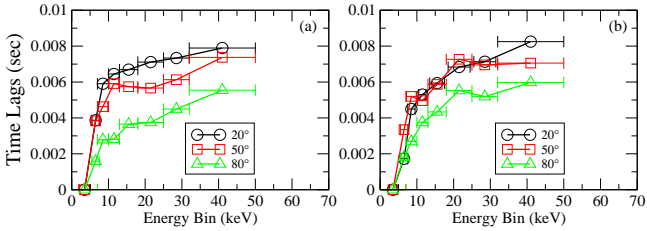}}
\caption{Time lag w.r.t $2-5$ keV is plotted with energy bins for shock locations 
at (a) $30r_g$ (a) and (b) $55r_g$ for three different inclination angles. For both 
Compton cloud sizes, lag magnitude decreases with inclination angle.}
\end{figure}

For high inclination angle sources, the gravitational lensing and disk reflection play major 
role. This is the reason for smooth increase of lag with shock location for high inclination 
angle sources presented in the panel (c) of Fig. 4. Panel (b) represents the medium inclination 
angle sources. Neither lensing nor the reflection mechanism dominates for this case. So, 
this is a region of combination of three contributions, namely, Comptonization, disk 
reflection and lensing. This is reflected in the nature of the curves. Another reason is that
the temperature and density inside the CENBOL is not monotonic, being highest at the centre
of the disk. Thus, emerging photons of various energies see different optical depths when inclination 
angle changes.

Panel 4(a) shows the variation for the case with very low inclination angle where focusing is much stronger 
than the disk reflection component. Curved geodesics are focussed in axial direction due to the
strong gravity of black hole. Here, we can see $t_{lag}$ increases with shock location. But, after 
a certain CENBOL size it starts to decrease. For Keplerian disk also, the inner region photons 
come earlier than that of the outer region as shown in Fig 7a when the view is  `pole on'. 
But, in presence of Comptonization, for smaller CENBOL size, the Comptonization time compensates 
the arrival lead of hard photons from the inner regions. But, when CENBOL size grows bigger, 
the relatively harder photons scattered from the top region of CENBOL will surely reach the 
observer before the photons from Keplerian disk as the inner edge of Keplerian disk is far away from 
the line of sight of the observer. This makes the change in slope shown in Fig. 4a.
Such an effect is only visible when the CENBOL is big. The light travel time increases 
the time delay between the near and far side of the
Keplerian disk with increasing inclination angle. But, the secondary photons and the photons
from inner regions remain almost at the similar time of arrival region for medium and high
inclination angle cases (compare Fig. 7(b) and 7(c)). This variation of time of arrival delay
between the photons originated from two sides causes smooth increase of time lag in Fig. 4c.

In Fig. 9, we plot the time lag variation with energy for different inclination angles
for a fixed CENBOL size: (a) for $30r_g$ and (b) for $55 r_g$.  
For the same accretion disk geometry, the magnitude of time lag decreases with 
increasing inclination angle. From Fig. 7, it is clear that an accretion disk
of diameter $60~r_g$ would cause $60~r_g/c$ time delay due to light crossing. So, in 
this current context, the diameter being $200~r_g$, this would cause $200~r_g/c$ 
($0.02~s$ or even more if secondary photons and disk reflection are considered) 
delay of arrival of the photons from the other side of the Keplerian disk. But, this 
contribution will reduce the time lag due to Comptonization if a maximum number 
of photons (for higher optical depth of the cloud) from the other side is intercepted 
by the CENBOL region. Thus, the lag magnitude decreases with inclination angle and 
the difference of time lag values (Fig. 9) for high and low inclination angle of a 
given CENBOL size are in agreement. 

In CCG17, the appearance 
of the Keplerian disk through the Compton cloud is shown for particular optical depth 
cases. When photons from the far side of the Keplerian disk gravitationally 
bend and pass through the CENBOL to reach the observer, the travel time of soft photons increases.
Since the time of the reference band has increased significantly by gravitational bending
the relative lag between two bands or lag magnitude has decreased for higher inclination angles.
Due to bending of light, it is possible to find negative lag for high inclination angle 
sources for a critical shock location $X_c$ and optical depth (see DC16 for further discussion). This 
happens when the shock strength is increased. These cases will be explored elsewhere.
  
\section{Comparison with Observational Results}

Nowak et al. (1999), showed that the time lag decreases with Fourier frequency 
obtained for various energy bands. The energy dependent time lag of Cyg X-1 also
shows that the harder photons arrive late. Altamirano \& Mendez, 2015, reported
the variation of phase lag of GX 339-4 and Cyg X-1 with intensity. The study of 
time/phase lag is given importance since it provides the information about the 
accretion flow dynamics. Similar behavior of time lags were also reported for
Supermassive Black Holes (see Kara et al. 2013 for details). Recently,
DC16, found the correlation between time lag and QPO centroid frequency for 
outbursting candidates such as XTE J1550-564 and GX 339-4 where they have shown the
magnitude of time lag is inversely proportional to the QPO frequencies.   
These basic natures of time lag are reported in literature. Here, 
in this section, we show the simulated results of correlation between QPO frequencies 
and time lags as obtained using TCAF solution.  

According to CM00, the shock location is inversely related to the QPO: larger the 
CENBOL size or shock location, smaller is the value of QPO centroid frequency. We consider 
the biggest ($65r_g$) CENBOL size is the first occurance. In an outburst source, it 
shrinks day by day to produce softer spectra. The evolution mechanism 
can be found in details in Debnath et al. (2014); Mondal et al. (2014); Molla et al. 
(2016); Jana et al. (2016); \& Chatterjee et al. (2016). We assume the shock is 
moving inwards with a constant velocity of $500~ cm/s$. The value of shock velocity 
is of the same order with that of the results obtained by Molla et al. (2016). 

\begin{figure}
\centering
\vbox{
\includegraphics[width=1.0\columnwidth]{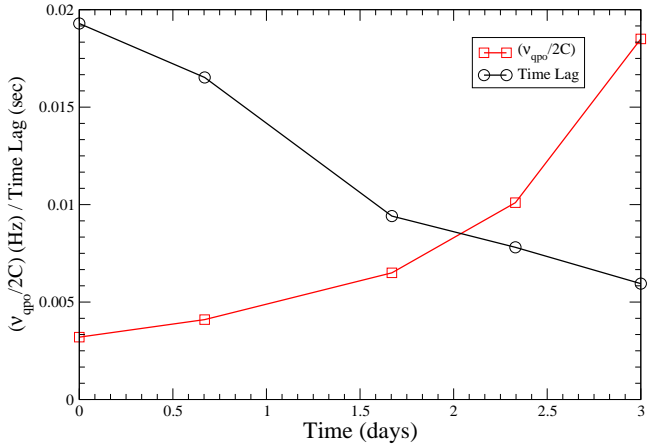}}
\caption{Time lag between $5-100$ and $0.1-5$ keV (solid-black-circle) photons 
is plotted for $50^{\circ}$ inclination angle along with calculated $\nu_{qpo}/2C$ 
Hz (solid-red-triangle). X-axis represents the day of evolution.}
\end{figure}

For a fixed inclination angle ($50^\circ$), we plot the time lag variation with the 
day along with ($\nu_{qpo}/2C$) where $C$ is a constant (CM00). From Fig. 6, we see 
that the time lag is maximum for lowest QPO occurance and it slowly decreases with 
increasing QPO frequency. This result is in complete agreement with the Fig. 4 of 
DC16 where they showed a similar variation for GX 339-4. Inclination angle of this 
source is reported as ($\sim 50^\circ$) (Zdziarski et al. 1998).

\section{Discussions and Conclusion}

From our simulations, we clearly see the time lag varies with the size of the CENBOL, 
acting here as the Compton cloud. The lag variation with energy of photons is also clearly 
present in both observation and our simulation. Explicit dependence of time lag with 
inclination angle can be understood from our work. For very low inclination objects, 
we have simulated results which shows soft lags for ($5-7$) keV and ($7-10$) keV 
energy bins. We speculate that this lag will be there for Hard and Hard Intermediate 
states where the CENBOL size is big enough. 

For high inclination angles, sources tend to have more hard photons since there is paucity of photons from the standard disk.
Earlier in the RXTE era, the energy dependent study of lag was limited up to 25 keV. 
Recently, ASTROSAT provided the data to study lag up to 50 keV (e.g., Jadav et 
al 2016; Mishra et al. 2016). Observationally, soft lags are mostly found for the 
high inclination angle sources such as, XTE J1550-564, GRS 1915+105, H1743-322, MAXI 
J1543-564 and XTE J1859+226 (see, DC16; Yadav et al. 2016; Mishra et al. 2017 and 
Eijnden et al. 2017 for details). In our picture, this effect is due to the 
down-scattering of outgoing photons at the regions immediately inside the CENBOL 
when the shock strength is high, especially near the equatorial plane. 
In a time independent model of CENBOL (following C85), 
this is not possible to show unless we carry out time dependent fluid dynamics 
results coupled with Monte-Carlo simulation which is very time consuming. The 
behavior of CENBOL in time dependent simulations are different than the thick disk 
geometry (for details see Molteni, Lanzafame, \& Chakrabarti, 1994 (MLC94); Ryu, 
Chakrabarti, \& Molteni, 1997 (RCM97), Giri \& Chakrabarti et al. 2010; Giri \& 
Chakrabarti 2012; Giri, Garain \& Chakrabarti 2015). The electron density inside 
Compton cloud is sharply high at the outer edge due to shock compression. 
The advection component opens up the cusp region and outflows are 
present in time evolution. Photons could suffer down-scattering at the 
post-shock region and this increases the soft photon travel time. Soft lags 
are mostly found for cases where the QPO frequencies are relatively higher and 
the system is mostly in Hard Intermediate State (HIS). In this state, the system 
is expected to have jet activity. The signature of jets for XTE J1550-564 in its 
1998 outburst has been reported in Wu et al. (2002). The lag behavior started to 
change the sign on MJD 51073 and the jet activity is increased rapidly at that time. 
So, the inclusion of jets and time dependent features of Compton cloud is expected 
to improve the results further. In this paper, our motto is to discuss the origin of time 
lag behavior with a very simplified TCAF solution. However, even with this simplified model,
we obtain a behavior similar to the observational results (Fig. 10). Time dependent aspects
of time lag behavior and its evolution with the spectral states are beyond the scope of the
present paper and will be discussed elsewhere.

\section{Acknowledgements}

The work of AC is supported by Ministry of Earth Science (MoES), India.
 


\end{document}